\documentclass[4p]{elsarticle}
\usepackage{graphicx}
\usepackage{amsmath,amssymb,bm,amsfonts}
\usepackage{slashed}

\newcommand{\ecm}{\mbox{$e$\,cm}}
\newcommand{\efm}{\mbox{$e$\,fm}}
\newcommand{\half}{\mbox{$\frac{1}{2}$}}
\allowdisplaybreaks[1]
\begin{document}

\begin{frontmatter}
\title{Electric dipole moments of the nucleon\\[0.2em] and light nuclei}

\author[Juelich]{Andreas Wirzba}

\address[Juelich]{Forschungszentrum J\"ulich, 
             Institut f\"ur Kernphysik,
             Institute for Advanced Simulation, 
         and J\"ulich Center for Hadron Physics,
             D-52425 J\"ulich, Germany}
\date{\today}

\begin{abstract}
The electric dipole moments of the nucleon and light ions are discussed and strategies for disentangling the
underlying sources of CP violation beyond the Kobayashi-Maskawa quark-mixing mechanism of the
Standard Model are indicated. Contribution to {\em ``45 years of nuclear theory at Stony Brook:
              a tribute to Gerald E. Brown''.}
\end{abstract}

\end{frontmatter}

\section{Prologue}
I came to Gerry Brown's group in 1982 as a visiting graduate student 
on a one-year scholarship of the {\em Studienstiftung des deutschen Volkes}
on recommendation of Achim Richter and Hans A.~Weidenm\"{u}ller. Gerry with his {\em big heart} 
readily integrated me in his group and
eased my way into the graduate school in Stony Brook. He immediately put me on a project about
pion-absorption on heavy nuclei  which he, Wolfram Weise and Hiroshi Toki had
been working on for some time. As I could correct some mistakes in the evaluation of the branching ratios, I earned---according to Gerry---my place as a co-author on a common paper which was published already in 1982. 
It was my highest cited paper for quite a while.  After extending this work to  
$^4$He together with the late Bernd Schwesinger, I then gradually entered the world of skyrmions and
Casimir calculations of the chiral bag which blossomed during my stay at NORDITA and the Niels-Bohr-Institute
from summer 1983 to 1984. Gerry moved with his `cloud' of students (which included Ulf Mei{\ss}ner and 
Dubravko Klabucar) to Copenhagen, where I  also met Ismail Zahed who was then a new postdoc hired by Gerry.

I  have always admired Gerry's intuition which enabled him, without nearly any mathematical apparatus, 
to grasp the essentials of  physics phenomena and to predict even  the correct sign and magnitude.
As far as I know, Gerry had never worked on electric dipole moments (EDMs), maybe because of his correct intuition that
a positive   EDM measurement of any subatomic particle will not 	materialize in his lifetime.
I am also not sure whether I will see one.  Well, as in any conference, there has to be a talk
which Gerry would be least interested in. I am afraid that it could be mine, however, unfortunately we cannot ask him for
his opinion any longer\,...\,.

\section{Motivation: matter-antimatter asymmetry in the universe}
No matter how much {\em matter} in comparison to {\em antimatter}  might have 
been created at the big bang itself, at the end of the
inflation epoch the  baryon--antibaryon (density) asymmetry must have  been diluted to a high precision: 
$n_{B}= n_{\bar B}$.
 However, about $3.8\cdot 10^5$ years later, when electrons and protons 
combined to form the first  hydrogen atoms such that the corresponding photons could `freeze' out from the evolution of
the universe, this asymmetry---weighted relative to the  photon density $n_\gamma$---acquired the following
value
\[ 
 \frac{n_B - n_{\bar B} }{n_\gamma}\Big |_{\text{CMB}} = (6.08 \pm 0.09)\cdot 10^{-10} \,.
\]
This was inferred from the cosmic microwave background (CMB) measurements by the 
COBE and WMAP satellite missions, where the displayed value is from a recent update~\cite{Bennett:2012zja}.
The above displayed number has to be compared with the prediction of the Standard Model (SM) of particle physics
 which is about 7 orders of magnitude less, $n_B/n_\gamma\big|_{\text{CMB}} \sim 10^{-18}$, where this value
 follows from the incorporation of 
 the determinant~\cite{Bigi_Sanda} of the Cabibbo-Kobayashi-Maskawa (CKM) quark-mixing matrix~\cite{Cabibbo:1963yz,Kobayashi:1973fv}
of the SM.

In 1967 the eminent Russian physicist Andrey Sakharov~\cite{sakharov} formulated three conditions for the dynamical
generation of net baryon number during the evolution of the universe:
\begin{enumerate}
\item There has to be a mechanism for the generation of baryon charge $B$ in order 
to depart from the initial value $B=0$ (after the
inflation epoch).
\item There should be C and CP violation to distinguish the rates of $B$ production from the   $\bar B$ production.
\item The dynamical generation had to take place during a stage of  non-equi\-librium, as 
otherwise the time-independence in the equilibrium phase would induce, under the assumption that CPT invariance
holds, CP invariance in the average,  such that also $\langle B \rangle =0$ holds in the average.
\end{enumerate}
Whereas $B$ violation, more precisely,  baryon plus lepton number violation $B+L$ can be accommodated by
the  Standard Model via the sphaleron mechanism at early temperatures \mbox{$\sim$ 1\,TeV}, 
the other two conditions cannot be met by the SM, since the
CP breaking by  the Kobayashi--Maskawa (KM) mechanism~\cite{Kobayashi:1973fv}
 of the SM is too small and since the SM at vanishing chemical potential
shows only a rapid cross over instead of a phase transition of first order. Therefore, the matter-antimatter asymmetry
together with the insufficient CP violation of the SM
is one of the few existing indicators that there might be
physics beyond the Standard Model (BSM physics).

\section{Electric dipole moment}
\subsection{Generalities}
How does the (permanent) electric dipole moment (EDM) fit into this?
From standard electrodynamics we know that the electric dipole moment is  a {\em polar}  spatial vector which measures
the permanent displacement of electric charges inside 
a given system.  As a polar vector it should change its sign under parity (P), but not under  time reversal (T).
Let us now assume that the system is a  massive subatomic particle  in its ground state. 
In its center-of-mass frame the only vector at our disposal is its spin which, however, is an {\em axial
vector} and therefore has the opposite transformation behavior under these discrete symmetry transformations as a polar vector.
 A subatomic particle can therefore only support a permanent electric dipole moment (vector)
\begin{equation} 
 \vec d = \sum_i e_i\vec r_i  \  \xrightarrow[\text{particles}]{\text{subatomic}} \  d\, \vec S/S
\end{equation}
with a non-vanishing coefficient $d$ 
if both P and T are violated. 
  Assuming that the CPT theorem holds, {\it i.e.} the validity of a local, hermitian, and relativistic field theory, 
  the  violation of  P and T  also implies   CP (and CT) violation.
  
  \subsection{Existence theorem for permanent electric dipole moments}
  We can summarize what was said above by the following theorem  which describes the existence of  
  permanent EDMs:
 {\em  \begin{quotation}
  \noindent Any non-vanishing coefficient $d$ 
   in the relation of the expectation values 
  \[
      \langle j^P | \vec d|  j^P \rangle = d \langle j^P| \vec J | j^P \rangle
  \]
  of  the  electric dipole moment operator $\vec d \equiv \int \vec r \rho(\vec r) d^3 r$
   and the total angular momentum $\vec J$ expressed in terms of 
   a stationary state $|j^P\rangle$ of a particle with at least one nonzero generalized `charge',
   nonzero total angular momentum (or spin) 
  $j$, nonzero mass, definite parity $P$ and no other degeneracy
  than its rotational one is  a signal
  for P and T violation and, because of the CPT theorem, for flavor-diagonal CP violation.
  \end{quotation}}
   Thus any nonzero measurement of an EDM of such a particle
might be interpreted as  ``a look through the rear window'' to the CP violation in the early universe.

 The above particle can be an {\em `elementary'} particle as a quark, charged lepton, $W^\pm$ boson, Dirac neutrino, etc., or
 a {\em `composite'} particle as a neutron, proton, nucleus, atom, molecule or even a solid body, as long as it meets
 the requirements stated in the above theorem.
 This might raise some questions~\cite{Lamoreaux}:
 
   {\em Isn't an elementary particle a point-particle without structure? How
 can such a particle be polarized and support an EDM?} 
 Well, we know that there are always vacuum polarizations and vertex corrections with rich short-distance structure. The gyromagnetic
 ratios $g$  minus 2 of an electron or  muon are also not exactly zero either, as they would be if the
 electron and muon were just Dirac point-particles.

 {\em What about the huge, measured EDMs of H$_2$O or NH$_3$ molecules which are of the order $10^{-8}\,\ecm$?}
 The ground states of these molecules at {\em nonzero} temperatures or {\em sufficiently strong} 
 electric fields are mixtures of
 at least two opposite parity states, such that they are not states of definite parity. The
  above theorem does not apply. 
  The measured nonzero EDMs  of
 water or ammonia molecules are therefore  neither signs of P nor T violation.
 Note, however, that the ground states of all known subatomic particles meet the condition of non-degeneracy.

{\em What about the induced EDM (i.e. about polarization)?}
While the coupling of the {\em permanent} EDM is {\em linear} to the electric field (linear Stark effect),
the coupling of the {\em induced} EDM, where the charges of the particle are   polarized by the electric field, is {\em quadratic} to the electric field (quadratic Stark effect).
The spatial vector necessary
to define the EDM  is in the induced case provided by the electric field  $\vec E$ itself, which, of course, 
is a polar vector, or by the spin multiplied by the projection of the electric field onto the spin 
$\vec S (\vec E\cdot \vec S)$.  Therefore the induced EDM  neither  signals P nor T violation. 

In fact, if the temperature or
electric field applied to the  above mentioned molecules ({\it e.g.}  H$_2$O or NH$_3$)
 are so small that  the ground state can be separated from the first excited
state of opposite parity, one first measures an induced EDM (quadratic
Stark effect). If the temperature or electric field are then increased, such that these two states cannot
be separated any longer, only then the linear Stark effect with
the measured huge permanent EDM takes  over in the molecular case. But in both cases there is neither P nor
T violation.

 \subsection{EDM  estimates, empirical windows and bounds} 
 In the following we will give a naive estimate for the size of the  permanent electric dipole moment $d_N$ of the nucleon~\cite{Lamoreaux}:
 
\noindent (i)   The size estimate of $d_N$ has to start with  
 the scale of  the CP and P conserving (magnetic) moment of the nucleon, namely with
 the nuclear magneton $\mu_N$ which scales as
 \begin{equation}
   \mu_N = \frac{e}{2 m_p} \sim 10^{-14} \,\ecm\,.
  \end{equation} 
  Furthermore, a nonzero EDM requires  P and CP violation:
 
 \noindent(ii)  The {\em empirical price} to pay for P violation can be estimated by, {\it e.g.} the product 
  \begin{equation}
      G_F \cdot F_\pi^2  \sim 10^{-7}
  \end{equation} 
 where $G_F \approx 1.166\cdot 10^{-5} \,\text{GeV}^{-2}$ is Fermi's constant  and
 $F_\pi \approx 92.2\,\text{MeV}$ is the axial decay constant of the pion, the order-parameter
 for the spontaneous breaking of chiral symmetry of Quantum Chromodynamics (QCD)
 at low energies~\cite{pdg:2012}. 

\noindent(iii)  The {\em empirical price} to pay for the CP violation can be estimated from, {\it e.g.}, the ratio of the amplitude moduli
 of $K_L^0$ to $K_S^0$ decays into
 two pions~\cite{pdg:2012}:
 \begin{equation}
    |\eta_{+-}| = \frac{| \mathcal{A}(K_L^0 \to\pi^+\pi^-)|}{|\mathcal{A}(K_S^0 \to \pi^+ \pi^-)|} = (2.232\pm 0.011)\cdot 10^{-3}
        \,.
 \end{equation}

 In summary, the magnitude of the nucleon EDM starts to become interesting at the scale
 \begin{equation}
   | d_N | \sim 10^{-7} \times 10^{-3} \times \mu_N \sim 10^{-24}\, \ecm
\end{equation} 
 or smaller.
 In the Standard Model (without QCD $\theta$ term), the CP violation follows from the KM mechanism which
 only generates a  nonzero CP violating phase if  the CKM quark-mixing matrix includes at least three quark generations.
 This KM-generated CP violation is therefore  flavor-violating, while  the EDMs
 are, by nature, flavor-diagonal. That means that the SM (without QCD $\theta$ term) has to pay the additional
 price of a factor $G_F F_\pi^2 \sim 10^{-7}$ to undo the flavor violation---in summary
 \begin{equation}
   | d_N^{\text{SM}} | \sim 10^{-7} \times 10^{-24}\, \ecm \sim 10^{-31}\,\ecm\,.
\end{equation} 
This agrees in magnitude with three-loop estimates of Refs.~\cite{Khriplovich:1985jr,Czarnecki:1997bu},  
the two-loop estimates of Refs.~\cite{Gavela:1981sk,Khriplovich:1981ca} which include a strong {\em penguin} diagram and the long-distance effect  of a pion loop
and even modern {\em loop-less} calculations~\cite{Mannel:2012hb,Mannel:2012qk} with charm-quark propagators.
 The electron EDM in the SM is even further suppressed by a factor $10^{-7}$, 
 namely 
 $|d_e^{\text{SM}}| \sim 10^{-38}\,\ecm$, which follows from the replacement of a gluon loop by a weak-interaction one~\cite{Pospelov:1991zt}.
 
 From  the above estimated numbers one can infer that an EDM of the neutron measured in the
 window
  \begin{equation}
      10^{-24} \ecm  > |d_N| \gtrsim 10^{-30} \ecm 
      \label{window}
 \end{equation}
will be a clear sign for new physics beyond the KM mechanism of Standard Model: either {\em strong CP violation}
by  a sufficiently large
QCD $\theta$ term or genuinely new physics, as, {\it e.g.}, supersymmetric models, multi-Higgs models, or left-right-symmetric models. 

In fact, the experimental bound on the neutron EDM, which decreased from the pioneering work of Smith, Purcell and 
Ramsey in the 1950s \cite{Ramsey} by six orders of magnitude to the present value
$|d_n| < 2.9 \cdot 10^{-26}\,\ecm$ by the Sussex/RAL/ILL group~\cite{Baker:2006ts}, already cuts by two orders of
magnitude into the new physics window, excluding already some simple and minimal variants of the above mentioned
BSM models.

The corresponding quantity
for the proton, $|d_p| < 7.9\cdot 10^{-25}\,\ecm$, is based on a theoretical calculation~\cite{Dmitriev:2003kb}
applied 
to input from the EDM bound for the diamagnetic ${}^{197}$Hg atom, 
$|d_{\text{Hg}}| < 3.1 \cdot 10^{-29}\,\ecm$~\cite{Griffith:2009zz}, while the same method would predict
for the neutron the bound  $|d_n| < 5.8 \cdot 10^{-26}\,\ecm$ which is only slightly bigger than the 
Sussex/RAL/ILL limit~\cite{Baker:2006ts}.  

 The bounds on the electron EDM  are again indirectly  inferred from
theoretical calculations~\cite{PhysRevA.78.010502,PhysRevA.84.052108,Skripnikov:2013}, where
this time  the input is either from paramagnetic atoms, {\it e.g.}  ${}^{205}$Tl  with
 $|d_{\text{TL}}| < 9.4 \cdot 10^{-25}\,\ecm$ \cite{Regan:2002ta}, or from polar molecules, as {\it e.g.}
YbF~\cite{Hudson:2011zz,Kara:2012ay} or ThO~\cite{Baron:2013eja}. The latter experiment
provided for the most recent and best bound on the electron EDM: $|d_e| < 8.7 \cdot 10^{-29}\,\ecm$.

It is common to all the experiments mentioned above that they refer only to charge-neutral particles, since these
particles
can be stored in a trap in the presence of (reversible) external electric fields  which are needed for the EDM measurement.  
To achieve the same for charged particles, the trap can be replaced by a storage ring.
In fact, as a byproduct of the $(g-2)_\mu$ measurement, there exist a weak bound on the
EDM of muon $|d_\mu|  < 1.8 \cdot 10^{-19}\,\ecm$~\cite{Bennett:2008dy}. 

In order to measure the
EDMs of the proton and deuteron (and maybe helion) the srEDM Collaboration~\cite{Farley:2003wt,Semertzidis:2003iq}  and the JEDI Collaboration~\cite{Lehrach,Pretz:2013us} suggested to
use storage rings  with radial electric fields such that the presence of the electric dipole moment of 
longitudinally polarized stored charged particles, which are initially locked to the particle momentum by
the frozen-spin method, can be measured  by the build-up of the vertical polarization (to the ring plane) 
by a polarimeter. In the case
of the proton, a purely electric ring is sufficient and highly desirable, since systematical
errors can be reduced by counterrotating beams which can circulate simultaneously in the absence of magnetic
fields. In the case of the deuteron and helion additional magnetic fields are necessary to apply the 
{\em frozen spin method},
since the anomalous magnetic moments of these particles have the opposite sign to the one of the proton.
Whereas the final aim of the two  proposals is to measure the EDMs of the proton and light ions to an uncertainty
of $\lesssim 10^{-29}\,\ecm$, the JEDI Collaboration suggested to start with a  {\em precursor experiment}  at the
existing  COSY ring in J{\"u}lich, a purely magnetic ring, which, for this purpose, 
is modified by  a Wien filter. The latter
introduces a bias into the horizontal polarization plane, such that  eventually an 
accumulation of the vertical polarization follows via the induced bias. Mainly because of the  small length of
the Wien filter,
the expected uncertainty which can be achieved in
the precursor experiment will be just $\sim 10^{-24}\,\ecm$.
  
\section{EDM sources beyond the KM mechanism}
\subsection{Dimension-four sources: strong CP violation }  
Let us study the second source in the SM to induce CP or rather P  and T breaking, the strong CP  violation
by the QCD $\theta$ angle.  Because of the topologically non-trivial nature of the QCD vacuum,
there exists an additional Lagrangian term of dimension four which respects all QCD symmetries (especially
the local color SU(3) symmetry), except the discrete   P and T ones:
\begin{equation}
  \mathcal{L}_{\text{QCD}} = \mathcal{L}_{\text{QCD}}^{\text{CP}} 
   + \theta \frac{g_s^2}{64 \pi^2} \epsilon^{\mu\nu\rho\sigma} G^a_{\mu\nu} G^a_{\rho\sigma}\,,
\end{equation}
where $G^a_{\mu\nu} = \partial_\mu A^a_\nu - \partial_\nu A^a_\mu +g_s f^{abc} A^b_\mu A^c_\nu$ is the
pertinent non-abelian field strength tensor and $f^{abc}$ the structure constant of SU(3).
With the help of  an axial U(1)$_A$ transformation of the quark fields, 
the parameter $\theta$ can be rotated from the above Lagrangian into the phase of the
 determinant of the quark-mass matrix $\mathcal{M}$:
\begin{equation}
 \dots +  \theta \frac{g_s^2}{64 \pi^2} \epsilon^{\mu\nu\rho\sigma} G^a_{\mu\nu} G^a_{\rho\sigma}
\ \xrightarrow{\text{\ U(1)}_A} \
  \dots - \bar \theta\, m_q^\ast \sum_{f=u,d} \bar q_f i\gamma_5 q_f\,,
  \label{theta_trans}
\end{equation}
where $m_q^\ast = m_u m_d/(m_u+m_d)$ is the reduced quark mass and where
\begin{equation}
  \bar \theta = \theta + \text{arg} \,\text{det}\mathcal{M}
 \end{equation} 
 is a further QCD parameter in addition to the quark masses.
Because of the coupling of $\bar\theta$ to the reduced quark mass, the $\theta$ term could be completely removed if
at least one quark had a vanishing mass which, however,  is empirically excluded~\cite{pdg:2012}.
The neutron EDM induced by the strong CP violation of the SM scales therefore as
\begin{equation}
  | d_n^{\bar\theta} | \sim \bar\theta  \cdot \frac{m_q^\ast}{m_s} \cdot \frac{e}{2 m_n} \sim \bar\theta \cdot 10^{-2} \cdot
  10^{-14} \,\ecm \sim \bar\theta\cdot 10^{-16}\,\ecm\,,
  \label{dn_theta_est}
\end{equation}
where the first term takes care of the reduced quark mass in terms of the strange quark mass $m_s$,  
the last removed scale,\footnote{If $m_q^\ast$ includes the strange quark mass, then $m_s$ in Eq.~(\ref{dn_theta_est}) has to be 
replaced by $\Lambda_{\text{QCD}}$.}
while the second term is the usual nuclear magneton. In naive dimensional analysis (NDA), $\bar\theta$ should
be of order one. The experimental bound on the neutron EDM, $|d_n| \lesssim 2.9 \cdot 10^{-26}\,\ecm$~\cite{Baker:2006ts}, however, limits this
parameter to
$
  | \bar \theta| \lesssim 10^{-10} 
$.
This tremendous deviation from NDA is called the {\em strong CP problem}.
On the other hand, the `new physics' window (\ref{window}) for  neutron EDM measurements  directly 
leads to the following window of  $\bar\theta$ values which would signal physics beyond the KM mechanism:
\begin{equation}
 10^{-10} \gtrsim  |\bar\theta| \gtrsim 10^{-14}\,.
 \label{window_theta}
 \end{equation}
However, as argued in Ref.~\cite{Kuzmin} these values are already too small to explain  the CP violation needed for
the
cosmic matter surplus, mainly because the chiral symmetry breaking scale  of QCD, 
$\Lambda_{\chi\text{SB}}\sim  1\,\text{GeV}$ is very small in comparison to the
electroweak-symmetry breaking scale 
$\Lambda_{\text{EWSB}}\sim 100\,\text{GeV}$.
  
 Thus  CP violating sources of higher dimension than four are needed to explain the baryon--antibaryon asymmetry
 in the universe.
  
  \subsection{Sources of dimension beyond four}
\begin{figure}[ht]
 \begin{center}
  \includegraphics[width=0.75\textwidth]{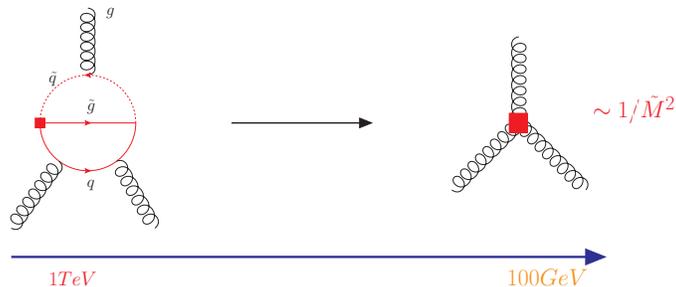}
  \caption{Example for the generation of an effective CP-violating three-gluon operator, the so-called Weinberg operator
  (see Refs.~\cite{Pospelov_review,Engel:2013lsa} for reviews and notations)
  from a SUSY two-loop process with a quark-gluino-squark coupling. The resulting operator is of dimension six as signalled by the suppression by the square of the BSM scale which here is called $\tilde M$. \label{fig:Smeft}}
  \end{center}
  \end{figure}

  The question is how to handle CP-violating sources which arise from physics beyond the SM, 
 {\it e.g.} from supersymmetric (SUSY), multi-Higgs, left-right-symmetric models, etc.
 The answer is a four-step procedure in the framework of effective field theory (EFT):

\noindent (1)  First, one picks a  BSM model (or class of such a  model) at a scale above $\Lambda_{BSM} > \mathcal{O}(m_t,m_{\text{Higgs}})$.
 Then all degrees of freedom beyond the BSM scale have to be integrated out, such that only SM degrees of freedom remain: namely quarks, gluons, Higgs, $Z$,  $W^\pm$.
 For that purpose one has to write down {\em all} interactions, even non-renormalizable ones with operators of dimension 
 six and higher,\footnote{Some of the interactions involve operators which naively seem to be only of dimension five. 
 However, the SM symmetries enforce the insertion of at least one Higgs field at high energies which transcribes
 to at least one (light) quark mass insertion at low energies.} for these {\em active} degrees of freedom
 that respect the SM and Lorentz symmetries. Of course one needs a power counter scheme to order
 the infinite number of interactions. Relics of the eliminated BSM physics are `remembered' by the low-energy
 constants (LECs) of the CP-violating contact terms of dimension six
  or higher. (See Fig.~\ref{fig:Smeft}.)
 
\begin{figure}[ht]
 \begin{center}
  \includegraphics[width=0.95\textwidth]{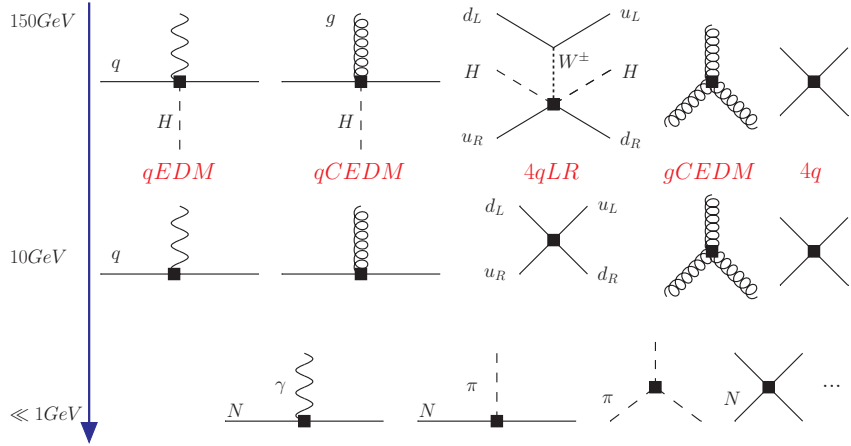}
  \caption{The four-step process (the second and third step are collapse to one step in the figure). Shown
  are the graphical representations of the quark EDM, the quark chromo\,EDM, the 4-quark left-right, the
  gluon chromo\,EDM, and the chiral symmetric 4-quark operators. All these operators mix to generate  the displayed
  CP-violating photon-nucleon, pion-nucleon, three-pion, and four-nucleon vertices, etc., see Ref.~\cite{Dekens:2013zca}.
  \label{fig:eftchpt}
  }
  \end{center} 
  \end{figure}

\noindent (2) In a second step all degrees of freedom beyond the electroweak (EW) scale are integrated out, such that
only the gluon and the five lightest quark degrees of freedom remain {\em active}. (See Fig.~\ref{fig:eftchpt}.)

\noindent(3)
 In a third step, the operators below $\Lambda_{\text{EW}}$  are scaled down by one-loop QCD 
 renormalization-group equations
 to the chiral symmetry breaking scale $\Lambda_{\chi}\sim 1\,\text{GeV}$, see {\it e.g.} Ref.~\cite{Dekens:2013zca}.
 The total number of  resulting independent operators is  nine  purely hadronic ones (including the $\bar\theta$ term) and three semi-leptonic ones. 
 Of the eight BSM operators of dimension six there are the isospin-conserving
 and isospin-breaking, respectively,  quark EDM  and quark chromo EDM operators 
 \begin{align} 
\text{(i)}\!:&   \ -\half e d_0 \bar  q\. i \sigma_{\mu\nu} \gamma_5q \,F^{\mu \nu} &\text{and}& \ \, &
\text{(ii)}\!:&  \ -\half e d_3 \bar q\tau_3 i\sigma^{\mu\nu} \gamma_5q \,F_{\mu \nu} \,, \\
\text{(iii)}\!:& \ -\half \tilde d_0 \bar q t^a i\sigma^{\mu\nu} \gamma_5 q \,G^a_{\mu \nu} &\text{and}& \ \, &
 \text{(iv)}\!:& \  -\half\tilde d_3 \bar q\tau_3 t^a i \sigma^{\mu\nu}\gamma_5 q \,G^a_{\mu \nu} , 
 \end{align}  
where the coefficients $d_0,d_3,\tilde d_0,\tilde d_3$ implicitly include a quark mass dependence to render these
operators to dimension six ones.
There is furthermore
 the left-right 4-quark operator which already at the EW scale
 explicitly breaks isospin and chiral symmetry, since it stems from an extension of the electroweak sector of the SM
 (a tiny coupling of the flavor-changing charged current to  two Higgs-bosons and a right-handed quark). 
 Because of the one-loop QCD renormalization-group evolution it splits
 into an operator with only color-singlet bilinears and one with only color-octet bilinears.
 The ratio of the prefactors, however, is fixed: $\nu_8/\nu_1 \approx 1.4/ 1.1$~\cite{Dekens:2013zca},  such that
 the resulting operators are not linearly independent:
 \begin{align}
 \text{(v)}\!: &\  \  \  \, i\nu_1 V_{ud} \!\left ( \bar u_R \gamma_\mu d_R\, \bar d_L \gamma^\mu u_L 
                                               -\bar d_R  \gamma_\mu u_R\, \bar u_L \gamma^\mu d_L \right) \\
          &  \ \ +   i\nu_8 V_{ud} \!\left ( \bar u_R \gamma_\mu \lambda^a d_R\, \bar d_L \gamma^\mu\lambda^a u_L 
                                               -\bar d_R  \gamma_\mu \lambda^a u_R\, \bar u_L \gamma^\mu\lambda^a d_L \right)  \,.                                       
 \end{align}
Moreover, there is
 the gluon chromo EDM operator
 \begin{eqnarray}
 \text{(vi)}\!: \ \frac{d_{\text{W}}}{6} f^{abc} \epsilon^{\mu\nu\alpha\beta} G^a_{\alpha\beta} G^b_{\mu\rho} G^{c\,\rho}_{\nu}\,
 \end{eqnarray}
 which is  isospin symmetric and chirally symmetric. It  is already of dimension six and 
 doesn't  include a quark mass insertion.
 Finally there are  two independent chirally and isospin symmetric  4-quark operators of dimension six, one consisting of color\,singlet quark bilinears, 
 the other consisting of
 color-octet quark bilinears: 
 \begin{align}
 \text{(vii)}\!: & \  \, \mu_1\! \left(\bar u u\, \bar d i\gamma_5 d + \bar u i \gamma_5 u \,\bar d d 
               - \bar d i\gamma_5 u\, \bar u d - \bar d u\,\bar u i\gamma_5 d \right)\,,\\
 \text{(viii)}\!:  & \ \,  \mu_8 \!
                 \left (\bar u \lambda^a u\, \bar d i\gamma_5\lambda^a\! d + \bar u i \gamma_5 \lambda^a u \,\bar d\lambda^a d - \bar d i\gamma_5 \lambda^a u\, \bar u \lambda^a d
          -  \bar d \lambda^a  u\, \bar u i\gamma_5 \lambda^a \!d   \right)\!.
 \end{align}                
 In fact, the last three operators, the gluon chromo EDM and the two chirally symmetric four-quark operators cannot
 be separated by hadronic methods and will  therefore be counted as one  operator class.
 
  As a caveat  it 
  should be noted that implicitly $m_s \gg m_u,m_d$ has been assumed. If the EDMs were also expressed in
  terms of the strange quark mass $m_s$, the number of  independent T and P violating operators of dimension six
   would have been larger.
 
 \noindent(4) To go to even lower scales in the final step,  non-perturbative techniques have to be applied.
 This can be {\it e.g.} lattice QCD calculations or the application of chiral perturbation theory, suitably amended.
 The latter contains the underlying symmetries including any explicit breaking and
 the Wigner-Weyl versus Nambu-Goldstone realization, and the Goldstone theorem
 (the vanishing of the coupling of any Goldstone boson to other Goldstone bosons or general matter fields in the
 chiral limit) as {\em `translation table'} between the `quark/gluon language' and the `hadronic language'.
 The appearing P and T violating hadronic operators can be collected in the following effective 
 Lagrangian~\cite{deVries2011b,deVries:2012ab}
 \begin{eqnarray}
  \mathcal{L}^{\text P\!\!\!\!/ T\!\!\!\!/} &=& - 2  d_0 \bar N S^\mu N v^\nu F_{\mu\nu}
                                                             - 2 d_1 \bar N\tau_3 S^\mu N v^\nu F_{\mu\nu} \nonumber \\
  && \mbox{}+g_0 \bar N \vec \tau \cdot \vec \pi N + g_1 \bar N \pi_3 N  -  \frac{ \Delta}{ 2F_\pi} \pi_3 (\vec \pi)^2   
  \nonumber \\
                                                     && \mbox{}+        C_1 \bar N N \partial_\mu (\bar N S^\mu N) 
                                                           + C_2 \bar N \vec \tau N \cdot \partial_\mu (\bar N S^\mu \vec \tau N)
 \label{Lag_had}
 \end{eqnarray}
 where the coefficients of the seven terms are fed, with different strength, respectively, 
 by the underlying 9 (actually only 7) dimension six operators (including the $\bar\theta$ term).
For instance, the term with the coefficient $ \Delta$ gets to leading order only contributions from the
left-right four-quark term, while the contributions of the quark EDM to the P and T violating pion-nucleon terms
with the coefficients $g_0$ and $g_1$ and the chirally symmetric four-quark terms with the
coefficients $C_1$ and $C_2$ are suppressed by the factor $\alpha_{\text{em}}/4\pi \sim 10^{-3}$ because
of the induced one-photon loop. 

The $\bar\theta$ term because of its inherent coupling via
the reduced quark mass to the flavor-neutral pseudoscalar quark sources, breaks the chiral symmetry, but keeps
the isospin one. The consequence is that the isospin-symmetric $g_0^\theta$ pion-nucleon term is the leading order term, whereas the isospin-breaking $g_1^\theta$ term is of subleading order. In fact, in the NDA estimate~\cite{BiraEmanuele} it is of order
N$^2$LO (next-to-next-to-leading order). In chiral perturbation theory the corresponding low-energy
constant  can be derived from a CP conserving, but isospin-breaking pion-nucleon term and
 $g_1^\theta$ is predicted to be only NLO suppressed relative to $g_0^\theta$~\cite{Jan_2013}. 

For the qCEDM case both $g_0$ and $g_1$ are predicted to of leading order (LO) and of the same strength,
while in the 4qLR case, which developed from an isospin-breaking operator at the BSM and EW scales, 
only $g_1$ is of leading order. Finally, for the gCEDM and the remaining 4-quark operators, which all 
are chirally symmetric, the pion-nucleon couplings $g_0$ and $g_1$, because of the Goldstone theorem, have to be reduced by an additional quark mass insertion, such that all terms (except $\Delta$) are of the same order in these
cases.

\section{The Hadronic level}
 Let us have a look at the nucleon EDMs in the $\bar\theta$ case. 
 According to the chiral arguments of Refs.~\cite{CDVW79,Pich:1991fq,ottnad} the
 nucleon EDM is driven by the photon coupling to the loop-pion of the nucleon self-energy diagram with
 one normal CP conserving p-wave coupling and one CP-violating but isospin-conserving $g_0$ coupling.
 Still, since the loop is log-divergent and require renormalization, there have to exist counter terms of the same order as the $g_0$ loop term with undetermined  coupling constants, namely 
 $d_0^\theta$ and $d_1^\theta$ of Eq.~(\ref{Lag_had}).\footnote{Similar results apply for all dimension 
 six sources~\cite{deVries2011b}.}  From  the chiral perturbation theory point of view, the nucleon EDMs by themselves have
 therefore no predictive power. As argued by Guo and Mei{\ss}ner~\cite{Guo12} 
 the  two  counter terms (which are also governing the SU(3) case) have to be either
 fitted by data which do not exist yet or by lattice QCD. However, lattice QCD 
 calculations for single-nucleon EDMs~\cite{Shintani1,Shintani2} still
 apply at too big pion masses such that rather large systematic errors are expected.  
 
 If, however, the CP-violating nucleon self-energy diagram is cut in such a way that there is tree-level pion-exchange
 between two nucleons
 with one CP violating vertex and one CP conserving one and a standard photon coupling to a proton 
 propagator~\cite{Flambaum:1984fb},
 then this CP-violating process is of leading-order and contact interactions are suppressed by at least two
 orders of magnitude. So chiral perturbation theory has predictive power for the two-nucleon components of
 the $\bar\theta$-induced deuteron and helion EDMs~\cite{deVries2011b,Jan_2013,J.Bsaisou}.

 
  
  
  \begin{table}[h]
\begin{center}
\begin{tabular}{|c|c|c|}   
\hline
Reference &  potential  
& result  \\
\hline
\hline
{Liu and Timmermans} \cite{LiuTimmermans}     & A$v_{18}$	
& $1.43\times 10^{-2}$\\
\hline
{Afnan and Gibson} \cite{Afnan:2010xd} & Reid\,93	
& $1.53\times 10^{-2}$\\
\hline
{Song et al. } \cite{Song:2012yh} & A$v_{18}$ 	
&$1.45\times 10^{-2}$ \\
\hline
{Bsaisou et al.} \cite{Jan_2013} & CD\,Bonn	
& $1.52\times 10^{-2}$\\
\hline
\end{tabular}
\caption{Table of the $g_1$ contribution to the deuteron EDM in units of
$g_1 \,g_{\pi NN}\,\efm$. \label{DeuteronTab}}
\end{center}
\end{table}

  \subsection{EDM of the deuteron at leading order}
 The case of the deuteron is special as it acts as an isospin filter. The deuteron ground state is a $^3S_1$ state
 (with a small $^3D_1$) admixture. After a pion exchange involving the leading $g_0$ vertex which conserves the
 total isospin, the intermediate CP-violating wave function has to be in a $^1P_1$ state. Because  the electric
 interaction with the proton charge is spin independent, the intermediate state 
 cannot return to  the $^3S_1$\,--\,$^3D_1$ ground state
 and the matrix element vanishes. By exactly the same argument the leading contributions from the $NN$ contact
interactions vanish in the deuteron.  Thus the  $g_1$ vertex, which is subleading in $\bar\theta$ case and which is
isospin-breaking, is active instead.
In Refs.~\cite{Jan_2013,J.Bsaisou} the entries in Table~\ref{DeuteronTab} were collected which agree with each other
to about 10\,\% accuracy. In  addition, in Ref.~\cite{J.Bsaisou} $g_1$-interaction results for the ($\bar\theta$ induced)
deuteron EDM (up to and including N$^2$LO corrections) and $g_{0,1}$-interaction results for 
the ($\bar\theta$ induced) helion and triton EDMs (up to and including NLO corrections) 
have been reported, where, for the first time, the
calculations were done consistently in chiral perturbation theory (ChPT), with both the CP violating operators 
and the wave functions of next-to-next-to-leading order  
given by ChPT. This allowed for an estimate of the
pertinent uncertainty of the nuclear calculation which turn out to be 11\,\% in the deuteron case and 20\,\% in
the helion/triton one and which are considerably smaller than the hadronic uncertainties of the LECs in the $\bar\theta$ case, namely only $20\,\%$ of the LEC uncertainty in the deuteron case and $60\,\%$ in the helion/triton one.
Note that nuclear uncertainties relevant for diamagnetic atom EDMs can be  several 
hundred percent~\cite{Engel:2013lsa}.



\section{Conclusions}
Measurements of hadronic EDMs are characteristically 
of low-energy nature. Thus the predictions have to be given in
the language of hadrons. The only reliable methods to do so are lattice QCD and chiral perturbation theory,
since they guarantee  inherent and systematical uncertainty estimates.

The EDMs of light nuclei provide independent information to the nucleon ones, they may be even larger,
and, moreover, simpler. The deuteron and helion EDMs provide for orthogonal results, because of the
isospin-filter property of the deuteron.

It could very well be that the first non-vanishing EDM result might be detected in a charge-neutral case, {\it e.g.}
for the neutron or a diamagnetic or paramagnetic atom or even a molecule.
However, the measurements of light ion EDMs will play a key role in disentangling the underlying sources of CP violation,
since they are the only systems where the nuclear calculation can be performed with sufficient control such that
the uncertainties of the hadronic low-energy constants of the CP violating terms are not swamped.
So to disentangle the underlying sources for the CP violation, at least the EDMs of the proton, neutron, deuteron
and helion have to be measured.  

In summary, hadronic EDMs play a key role in hunting for new sources of CP violation.
They may be relevant for the observed baryon asymmetry of the universe (BAU).
However, there is no theorem which {\em directly} links the BAU with the EDMs. 
There can be leptogenesis instead of baryogenesis.
Remember also
that in the $\bar\theta$ scenario there may be sizable EDMs, without, however, enough strength for inducing the
observed BAU. Moreover, there are no smoking guns so far. Nevertheless, even the lower bounds of EDMs did 
provide  serious constraints for flavor-diagonal CP violating processes and CP violating sources beyond the SM and
the KM mechanism in the past and will do so, if just the bounds will be improved, in the future.

 \subsection*{Acknowledgements}

I would like to thank  Dima Kharzeev, Tom Kuo, Edward Shuryak and 
Ismail Zahed for organizing this wonderful meeting. I would also like
to thank all my collaborators, especially Jan Bsaisou, Christoph Hanhart, Ulf Mei{\ss}ner, Andreas Nogga, 
Kolya Nikolaev, Werner
Bernreuther, Wouter Dekens, and Jordy de Vries for sharing their insights into the topics presented here.

\newpage
\bibliography{bibliography}
\bibliographystyle{h-physrev3}

\end{document}